\documentclass[conference,a4paper]{IEEEtran}
\IEEEoverridecommandlockouts

\usepackage[hidelinks]{hyperref}
\usepackage[cmex10]{amsmath}
\usepackage{amssymb,amsfonts}
\usepackage{dblfloatfix}

\usepackage[ruled,vlined]{algorithm2e}
\usepackage{graphicx}
\graphicspath{{Figures/PDF/}{Figures/PNG/}}

\usepackage{booktabs}
\usepackage{siunitx}
\usepackage[numbers,compress]{natbib}
\usepackage{texnames}
\usepackage{bm,bbm}
\usepackage{orcidlink}
\usepackage{multirow}
\usepackage{microtype}
\usepackage{float}

\usepackage{subcaption}

\begin{document}

\title{\uppercase{Neural Posterior Estimation of Terrain Parameters \\from Radar Sounder Data}
\thanks{This work was supported by Italian Space Agency under Grant n. 2022-23-HH.1-2023 (CUP: F63C22000650005) and Grant n. 2023-6-HH.0 (CUP: F83C23000070005).
This work has been accepted at \href{https://2026.ieeeigarss.org}{IGARSS~2026} and might be submitted to IEEE for possible publication. Copyright may be transferred without notice, after which this version may no longer be accessible. 
}
}

\author{
    \IEEEauthorblockN{
        Jordy~Dal~Corso\,\orcidlink{0009-0004-3671-2402}\IEEEauthorrefmark{1},
        Annalena~Kofler\,\orcidlink{0009-0008-5938-6215}\IEEEauthorrefmark{2}\IEEEauthorrefmark{3},
        Marco~Cortellazzi\,\orcidlink{0009-0006-4145-0122}\IEEEauthorrefmark{1},\\
        Lorenzo~Bruzzone\,\orcidlink{0000-0002-6036-459X}\IEEEauthorrefmark{1},\,\IEEEmembership{Fellow~IEEE},
        Bernhard~Schölkopf\,\orcidlink{0000-0002-8177-0925}\IEEEauthorrefmark{2}\IEEEauthorrefmark{4},
    }
        
    \IEEEauthorblockA{\small
        \IEEEauthorrefmark{1}University of Trento, Trento, Italy\\
        \IEEEauthorrefmark{2}Max Planck Institute for Intelligent Systems, Tübingen, Germany\\
        \IEEEauthorrefmark{3}Max Planck Institute for Gravitational Physics (Albert Einstein Institute), Potsdam, Germany\\
        \IEEEauthorrefmark{4}ELLIS Institute Tübingen, Tübingen, Germany
    }
}

\maketitle
\begin{abstract}
	Radar sounders are electromagnetic instruments that can probe deep into the subsurface of Earth and other planetary bodies by processing the echo of transmitted radar waves.
    Conventional approaches for analyzing such data rely on approximate assumptions and often produce point estimates that ignore parameter correlations as well as galactic and measurement noise.
    We propose a simulation-based inference approach to terrain parameter inversion from radar sounder data, where synthetic observations from a GPU-based simulator are used to train a neural network-based density estimator for neural posterior estimation~(NPE).
    By explicitly conditioning on reference surface assumptions, the proposed framework allows systematic evaluation of posterior robustness to reference surface variability.
    We demonstrate that our NPE model is well calibrated on simulated data and transferable to real Mars radar profiles, where we analyze terrain parameters using literature-informed reference values.
\end{abstract}


\section{Introduction} \label{sec:intro}
\begin{figure*}[ht]
    \centering
    \includegraphics[width=\linewidth]{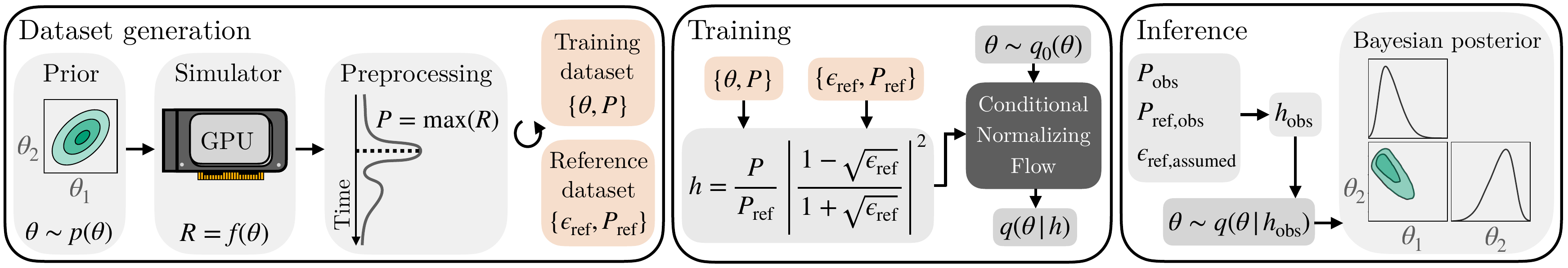}
    \caption{Overview of the NPE framework: During data generation, prior samples are passed through the simulator to extract peak power for both the training and reference datasets. 
    Random samples from both datasets yield the relative power~$h$, which is provided to the conditional density estimator which maps a base distribution~$q_0(\theta)$ to the posterior~$q(\theta \mid h)$. 
    For inference, observed powers~$P_{\mathrm{obs}}$ and~$P_{\mathrm{ref, obs}}$ are combined with an assumed $\epsilon_\mathrm{ref}$ to sample from the NPE model, resulting in the Bayesian posterior.
    \looseness=-1
    }
    \label{fig:npe_overview}
    \vspace{-1em}
\end{figure*}

Radar sounders~(RS) have enabled major advances in planetary science by revealing subsurface interfaces, buried structures, and ice deposits, thereby constraining the geological evolution of bodies such as the Moon~\cite{ono2010lrs, ono2009lunarss} and Mars~\cite{putzig2024science, grima2012quantitative}. 
These instruments transmit electromagnetic waves toward a planetary surface and analyze the returned echoes to probe the subsurface. 
The success of this technique has motivated the inclusion of RS in the payload of missions toward other planetary bodies, including the Radar for Icy Moon Exploration~(RIME) instrument aboard ESA’s JUICE mission~\cite{rime} and the Subsurface Radar Sounder~(SRS) on ESA’s Envision mission to Venus~\cite{bruzzone2020envision}. \looseness=-1

RS data are acquired in the form of rangelines, which consist of time series of echo samples recorded by the instrument. 
Each rangeline encodes information on the surface and the subsurface through intensity peaks associated with dielectric discontinuities.
As the instrument operates while moving along the azimuth direction, consecutive rangelines are collected and arranged into radargrams, which provide cross-sectional representations of the probed subsurface.
RS data are interpreted using both qualitative analysis, based on visual inspection of radargrams, and quantitative approaches relying on statistical parameter inversion techniques. 

Several approaches have been proposed to infer surface and subsurface properties from RS data, such as roughness, layer thickness, and material properties~\cite{plaut2009radar,mouginot2012dielectric}. 
Parameter inversion can also be performed using numerical simulations, in which echoes from a hypothesized terrain are modeled and directly compared with observations, either through one-to-one matching or systematic search strategies~\cite{Gerekos2018MRS,mastrogiuseppe2016radar}.
Despite the diversity of existing inversion approaches, none provides a comprehensive solution. 
Statistical methods are limited by simplified scattering models and typically yield point estimates, while simulation techniques that repeatedly compare measurements and simulations are computationally expensive and often restricted to localized analyses. 
Accurate information about the surface of planetary bodies can only be extracted with uncertainty-aware methods, motivating the adoption of a Bayesian inversion framework capable of incorporating prior information and delivering full-posterior parameter estimates.

Recent advances in RS simulators~\cite{Gerekos2018MRS,sbalchiero2023dictionary} and GPU acceleration have made large-scale, high-fidelity simulations feasible. 
These improvements enable the use of simulation-based inference (SBI) techniques~\cite{cranmer2020frontier}, specifically neural posterior estimation (NPE)~\citep{papamakarios2016, lueckmann2017, greenberg2019}. 
Unlike traditional Bayesian methods that require an explicit and often intractable likelihood function, NPE leverages synthetic datasets to learn the posterior densities.

In this work, we leverage these developments to introduce an NPE-based framework for RS data terrain parameter inversion, training an estimator on synthetic datasets generated by a GPU-based simulator to retrieve surface parameters. 
This approach overcomes the limitations of previous methods by offering amortized inference: computational costs are shifted to an offline training phase, enabling fast, scalable inversion and uncertainty quantification via a single forward pass at inference time. 
We explicitly address uncalibrated RS data by conditioning the inversion on the reference-zone dielectric constant and validate the framework on a notable region of Mars using SHARAD observations, demonstrating consistency with existing literature while providing full probabilistic parameter estimates. \looseness=-1

\section{Methodology} \label{sec:methodology}
\subsection{Neural posterior estimation}
NPE~is a SBI~technique focused on approximating the full Bayesian posterior distribution~$p(\theta|x)$ over the parameters~$\theta$ given data~$x$ with a conditional density estimator~$q(\theta|x)$~\citep{papamakarios2016, lueckmann2017, greenberg2019}.
To train such a density estimator, simulated data pairs $\{\theta, x\}$ are generated where samples are first drawn from the prior over the parameters of interest~$\theta \sim p(\theta)$ and then evaluated with an accurate, stochastic simulator~$x = f(\theta)$ (see Figure~\ref{fig:npe_overview}).

Different conditional density estimators can be employed for $q(\theta|x)$, with the most prominent being discrete and continuous normalizing flows~\cite{papamakarios2021normalizing, liu2022, lipman2023flow, albergo2023building, albergo2023stochastic, wildberger2023fmpe}.
Since discrete flows enable faster sampling and density evaluation than their continuous counterpart~\cite{wildberger2023fmpe}, we focus on them in this work.

The deviation of the density estimator~$q(\theta|x)$ from the true posterior can be quantified by the negative log-likelihood loss 
\begin{equation}
    \mathcal{L}_\mathrm{NPE} \approx - \mathbb{E}_{\theta \sim p(\theta), x \sim p(x|\theta)} \left[ q(\theta | x) \right]~.
\end{equation}
Once trained, samples from the posterior~$q(\theta | x_\mathrm{obs})$ can be obtained rapidly for any observation~$x_\mathrm{obs}$ consistent with the prior assumptions and the employed simulator, without the need for additional simulations.
As a result, the cost of training the density estimator can be amortized at inference time by re-using the same model for multiple measurements.

In our setting, observations $x$ consist of RS rangelines, and we aim to infer terrain parameters $\theta$ of the corresponding areas.
Since inference on $\theta$ is ultimately required for a large number of real RS~acquisitions, an amortized inference procedure is preferred over sequential NPE~methods.
For further details, we refer the reader to excellent overviews like~\cite{deistler2025simulationbasedinferencepracticalguide}.

\subsection{Simulation and dataset generation}
We generate pairs $\{\theta, x\}$ via an RS simulator that emulates the terrain backscattering process given terrain parameters $\theta$, illustrated in Figure~\ref{fig:npe_overview} on the left.
We define $x\equiv P$ as the peak power value extracted from a simulated rangeline, and the terrain parameters as the triplet $\theta\equiv(\epsilon,\sigma,m)$. Here, $\epsilon$ is the dielectric constant of the near-surface medium, $\sigma$ and $m$ are the standard deviation (RMS) of heights and slopes within the radar footprint, respectively. 
These parameters have been used in previous RS~studies on Mars~\cite{grima2012quantitative} and provide an effective low-dimensional description of surface properties~\cite{ulaby}.
While the proposed SBI~framework is general and applicable to any RS~simulator, we develop a dedicated simulation pipeline to retain full control over data generation.

We implement a GPU-based ray-tracing simulator inspired by approaches in \cite{berquin2015computing, Gerekos2018MRS} to generate synthetic rangelines. 
We define the simulator as a function $f: \mathbb{R}^3 \rightarrow \mathbb{R}^{n_\mathrm{s}}$ that maps $\theta$ to a simulated rangeline $R$, where $n_\mathrm{s}$ is the number of echo samples determined by the RS sampling frequency.
Radar settings (e.g. central frequency $f_c$ and wavelength $\lambda$) are treated as hyper-parameters and are not included as inputs.
We constrain the footprint size $D$ to the first Fresnel zone, and the size $d_x$ of the squared facets composing the surface to $\lambda/10$ (to apply the constant phase approximation~(CPA)~\cite{berquin2015computing}).

For each simulation, we generate the illuminated surface mesh as a 2D~Gaussian Random Field~(GRF) with given RMS height~$\sigma$ and slope~$m$.
The backscattered field from each facet is computed via the Stratton-Chu integral~\cite{berquin2015computing} with the phase term computed using the CPA. 
The increase in computational cost induced by the small facets is mitigated through batch processing on GPUs.
The total received signal is then constructed by summing the backscattered contributions~$E_i$ from all facets, each delayed by the two-way travel time $\tau = 2h_i/c$, where $h_i$ denotes the distance of facet $i$ from the radar. 
The resulting rangeline is given by:
\begin{equation}
R(t) = E_\mathrm{tot}(t) = n(t)+\sum_iE_i(t),
\end{equation}
where we added galactic noise $n(t)$ modeled as a Gaussian process with PSD~$S_\mathrm{galaxy}(f_n) = f_n^{-\alpha}$, $\alpha=2.5$ \cite{dulk2001calibration}. We finally extract the peak power as $P=\max_t |R(t)|^2$.

\begin{figure*}[ht]
    \centering
    \includegraphics[width=1\linewidth]{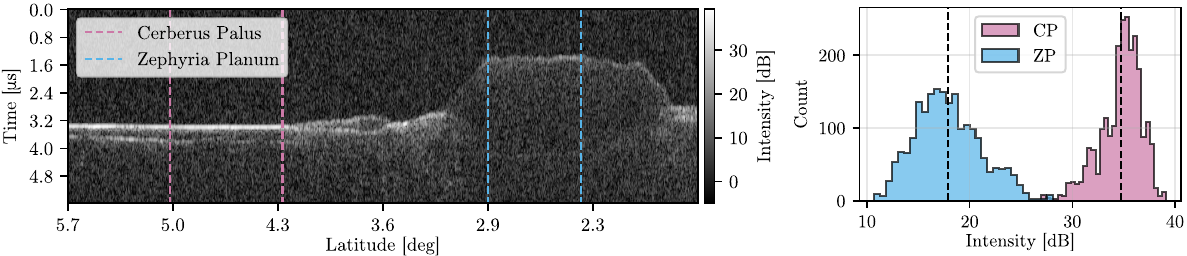}
    \caption{SHARAD Radargram 186301. We highlight in light pink and blue the CP and ZP~areas, respectively. On the right, we show the distribution of $P$ in rangelines within the regions and highlight their mean with dashed lines.}
    \label{fig:roi}
    \vspace{-1em}
\end{figure*}

\subsection{Conditioning on the reference surface} \label{sec:conditioning}
Orbital RS~data corresponds to uncalibrated power values, as calibration is generally unfeasible on ground.
To properly interpret such data, experts usually calibrate it against a reference zone (RZ) corresponding to a flat surface with assumed dielectric constant $\epsilon_\mathrm{ref}$, such as ice-covered regions~\cite{grima2012quantitative, castaldo2017global}.
Crucially, the $\epsilon_\mathrm{ref}$ assumed for the RZ may not correspond to the true one, impacting the inversion of terrain parameters.
Thus, inversion results based on RS~data depend on the assumption about~$\epsilon_\mathrm{ref}$.
To reflect this within the NPE~framework, we (1)~generate additional simulations of the reference surface, resulting in the reference data set~$\mathcal{D}_\mathrm{ref}$, (2)~normalize the peak power~$P$ from $\mathcal{D}$ by $P_\mathrm{ref}$, and (3)~include the assumption on~$\epsilon_\mathrm{ref}$ via:
\looseness=-1
\begin{equation} \label{eq:h}
    h = \frac{P}{P_\mathrm{ref}}\left|\frac{1-\sqrt{\epsilon_\mathrm{ref}}}{1+\sqrt{\epsilon_\mathrm{ref}}}\right|^2~.
\end{equation}
Equation~\ref{eq:h} is based on the ratio between the received power~$P$ from a~GRF and $P_\mathrm{ref}$ from a flat terrain with known $\epsilon_\mathrm{ref}~$\cite{haynes2018geometric, haynes2020surface}.
Since we provide~$h$ as an input to the density estimator~$q(\theta|h)$, it is not only conditioned on the relative power, but also on the assumption about~$\epsilon_\mathrm{ref}$, where one could write explicitly:~$q(\theta| h(P, P_\mathrm{ref}, \epsilon_\mathrm{ref}))$.
This procedure is illustrated in the center of Figure~\ref{fig:npe_overview}.

At inference time, we obtain samples from the density estimator~$q(\theta | h)$ for the measured peak powers~$P$ and~$P_\mathrm{ref}$, as well as an assumption about~$\epsilon_\mathrm{ref}$ which is shown on the right of Figure~\ref{fig:npe_overview}.
Since NPE allows rapid inference without additional simulator calls, it is possible to investigate how the posterior distribution is affected by different assumptions of~$\epsilon_\mathrm{ref}$.

\section{Experiments} \label{sec:experiments}
\begin{figure*}[ht]
    \centering
    \begin{subfigure}[b]{0.48\linewidth}
        \centering
        \includegraphics[width=0.95\linewidth]{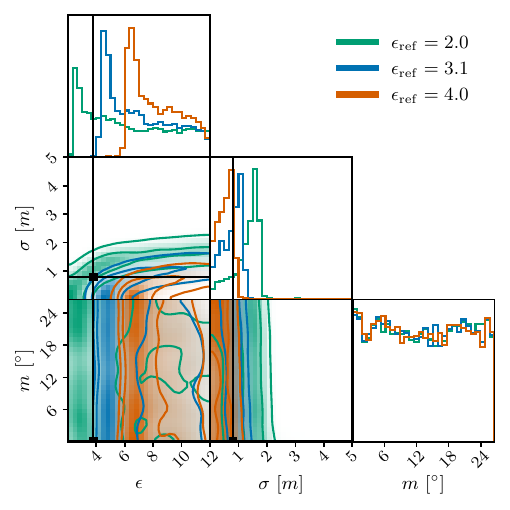}
        \vspace{-0.5em}
        \caption{CP posterior.}
        \label{fig:res-plain}
    \end{subfigure}
    \hfill
    \begin{subfigure}[b]{0.48\linewidth}
    \centering
    \includegraphics[width=0.95\linewidth]{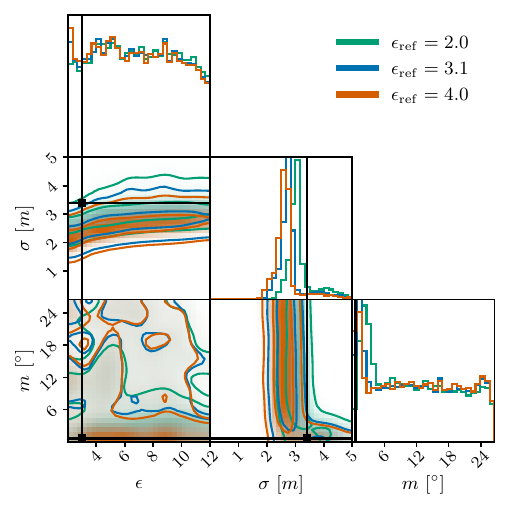}
    \vspace{-0.5em}

    \caption{ZP posterior.}
        \label{fig:res-hill}
    \end{subfigure}
    \caption{
    Inferred posterior distributions for CP and ZP under varying $\epsilon_\mathrm{ref}$. 
    Black lines illustrate values from \cite{alberti2012MarsPermittivity} and MOLA (see Table~\ref{tab:terrain_parameters}).
    Roughness from MOLA and $\epsilon$ values from~\cite{alberti2012MarsPermittivity} are within the posterior estimates on CP.
    }
    \label{fig:res-combined}
    \vspace{-1em}
\end{figure*}

\subsection{Dataset generation and training}
We generate two datasets with the simulator~$f$ to train and validate the posterior estimator.
The primary dataset $\mathcal{D}$ ($n_\mathcal{D}=10\,\mathrm{k}$) consists of peak powers~$P_\mathrm{peak}$ with parameters drawn from the prior.
Specifically, we sample $\theta_i \sim p(\theta)$ and simulate $P_i$ to form $\mathcal{D} = \{(P_i,\theta_i)\}_{i=1}^{n_\mathcal{D}}$.
We adopt uninformative priors to assess whether the inferred posteriors concentrate on physically plausible values: 
$p(\epsilon) = \mathcal{U}[2, 12]$, spanning ice to dense rock; $p(\sigma) = \mathcal{U}[0, 5]$, covering flat and moderately rough terrain, and $p(\sigma) = \mathcal{U}[0, 0.5]$, corresponding to  RMS slopes below $26^\circ$ as a conservative upper bound. 
The second dataset $\mathcal{D}_\mathrm{ref}$ ($n_\mathrm{ref}=2\,\mathrm{k}$), contains reference peak powers~$P_\mathrm{ref}$ (Section~\ref{sec:conditioning}) from a flat terrain where the reference permittivity~$\epsilon_\mathrm{ref}$ is sampled uniformly over~$[2, 4]$ to cover the values presented in~\cite{grima2012quantitative}. 
Training and validation datasets $\mathcal{D}_\mathrm{train}$ and $\mathcal{D}_\mathrm{val}$ are constructed by drawing $n_\mathrm{train}=80\,\mathrm{k}$ and $n_\mathrm{val}=20\,\mathrm{k}$ non-overlapping samples from the Cartesian product $\mathcal{D} \times \mathcal{D}_\mathrm{ref}$ and computing the corresponding $h$~values (Equation~\ref{eq:h}). \looseness=-1

For the architecture of the density estimator, we employ a coupling flow with rational-quadratic spline transformations~\cite{durkan2019neural} implemented using the \texttt{sbi} library~\cite{boelts2025sbi}. 
The flow consists of 5~subsequent transformations where the parameters of each transformation are predicted by a fully-connected network with a single hidden layer of 64~neurons. 
Training is performed for maximally 50~epochs using the Adam optimizer, with a batch size of~1024 and a fixed learning rate of~$10^{-3}$. 
To prevent overfitting, we include early stopping, and we select the model wit the lowest validation loss for inference.

\subsection{Validating the NPE model}
To validate the trained NPE density estimator, we perform simulation-based calibration~(SBC)~\cite{cook2006sbc, talts2018validating}. 
SBC evaluates whether the ranks of ground-truth parameters under posterior samples follow a uniform distribution \cite{talts2018validating}.
We assess calibration using (1)~Kolmogorov--Smirnov~(KS) tests on the rank distributions, (2)~classifier two-sample tests~(C2STs)~\cite{lopez2016revisiting}, which assesses whether the ranks and a uniform distribution can be distinguished by a binary classifier, and (3)~C2STs comparing the data-averaged posterior to the prior. 
The KS~tests result in high $p$-values for all parameters (median 0.85), consistent with uniformly distributed ranks. 
The C2ST accuracies are close to $0.5$, specifically $(0.58, 0.6, 0.57)$ for~(2) and $(0.50, 0.51, 0.54)$ for~(3) (ordered as $(\epsilon, \sigma, m)$).
This indicates that the distribution of ranks cannot be distinguished from a uniform distribution and the NPE model is well calibrated.

\subsection{Case Study: Medusa Fossae Formation}
When applying the NPE model to real data, we aim to (i)~verify that inferred terrain parameters are consistent with scattering laws and (ii)~asses the sensitivity of posterior estimates to the reference parameter~$\epsilon_{\mathrm{ref}}$. 
We evaluate the trained density estimator on SHARAD acquisitions over the Medusae Fossae Formation~(MFF) on Mars, a region exhibiting a geological dichotomy between the rough southern highlands and the younger northern volcanic plains. 
Our analysis focuses on echoes from radargram~186301, specifically Cerberus Palus~(CP) and Zephyria Planum~(ZP) (Figure~\ref{fig:roi}). 
Literature-based estimates for $\epsilon$ and roughness parameters from MOLA are summarized in Table~\ref{tab:terrain_parameters}. \looseness=-1

\begin{table}[ht]
    \centering
    \caption{\small Summary of values from \cite{alberti2012MarsPermittivity} and MOLA for CP and ZP.}
    \label{tab:terrain_parameters}
    \begin{tabular}{
        c c c S[table-format=3.0] c
    }
        \toprule
        Parameter & Region & Value & {Baseline [m]} & Source \\
        \midrule
        \multirow{2}{*}{$\epsilon$}
          & CP & $3.8 \pm 0.2$ & {--}  & \cite{alberti2012MarsPermittivity} \\
          & ZP & $3.0 \pm 0.5$ & {--}  & \cite{alberti2012MarsPermittivity} \\
        \midrule
        \multirow{2}{*}{$\sigma$}
          & CP & 0.8~ & 463 & MOLA \\
          & ZP & 3.4~ & 463 & MOLA \\
        \midrule
        \multirow{2}{*}{$m$}
          & CP & 0.1\si{\degree}  & 463 & MOLA \\
          & ZP & 0.6\si{\degree} & 463 & MOLA \\
        \bottomrule
    \end{tabular}
    \vspace{-1.5em}
\end{table}

For each region, we compute $P_\mathrm{peak}$ as the average of the maximum return per rangeline, yielding $P_\mathrm{CP} = \SI{34.73}{\decibel}$ and $P_\mathrm{ZP} = \SI{17.11}{\decibel}$. 
These values agree with~\cite{sbalchiero2023dictionary}, which reported a power difference of $\SI{16.5\pm 4.6}{\decibel}$ between the two regions.
We obtain~$P_\mathrm{ref}$ from a flat RZ located at $171^\circ\mathrm{E}, -84^\circ\mathrm{S}$. 
From SHARAD radargram \texttt{0178301}, acquired at altitude $r_\mathrm{ref}=250\,\mathrm{km}$, we measure $P_\mathrm{ref} = \SI{32.62}{\decibel}$ and rescale it to nominal altitude $r= \SI{300}{\kilo \meter}$ used in simulations via the linear correction factor $(r/r_\mathrm{ref})$~\cite{grima2012quantitative}.
In the following, we consider three reference permittivities~\mbox{$\epsilon_\mathrm{ref} \in \{2.0,\, 3.1,\, 4.0\}$}.

\subsection{NPE results on MFF}
To interpret the posterior distribution for CP (Figure~\ref{fig:res-plain}), we note that the measured power value~$P_\mathrm{CP}$ is close to the RZ~$P_\mathrm{ref}$ ($\Delta P= 2.11\,\mathrm{dB}$).
This implies either similar surface properties between CP and the~RZ or compensating combinations of terrain parameters, since distinct parameter sets can yield the same received power.
Assuming a flat reference surface, matching this power requires CP to be relatively smooth with a higher dielectric constant~$\epsilon_\mathrm{CP}$.
This trade-off is reflected in the $(\epsilon, \sigma)$ correlation where increasing roughness necessitates higher $\epsilon$ to offset roughness-induced power loss.
Increasing $\epsilon_\mathrm{ref}$ systematically shifts the inferred posterior toward higher $\epsilon$ and lower $\sigma$.
At larger $\sigma$, $\epsilon$ becomes weakly constrained, consistent with previous findings~\cite{grima2012quantitative}.
The posterior for the RMS height concentrates near the lower bound of the prior, around $\SI{1}{\meter}$, indicating a comparably smooth surface at CP. 
In contrast, the posterior marginal of the slope parameter~$m$ follows the prior, suggesting that the data are uninformative for this parameter. 
This is expected given the short baseline $d_x=\SI{1.5}{\meter}$ relative to the wavelength $\lambda=\SI{15}{\meter}$, which limits sensitivity to slope variations.
Results on~ZP (Figure~\ref{fig:res-hill}) are consistent with previous literature~\cite{grima2012quantitative}. 
The low measured $P_\mathrm{ZP}$ can be attributed to either low permittivity~$\epsilon$ or high roughness.
Given the inferred $\sigma$ of approximately $\SI{3}{\meter}$, the posterior for~$\epsilon$ remains uninformative and spans the prior range. 
Conversely, the posterior for $m$ favors lower values while still covering the full prior, suggesting that small $m$ may coexist with high $\sigma$, as in a tilted but locally smooth surface. 
As $\sigma$ increases, the influence of $\epsilon_\mathrm{ref}$ on the inferred posteriors diminishes. \looseness=-1

\section{Conclusions} \label{sec:conclusions}
In this work, we presented a simulation-based inference framework based on neural posterior estimation for terrain parameter inversion from radar sounder data, yielding full Bayesian posterior distributions. 
The approach leverages NPE trained on synthetic data generated by a custom GPU-based RS simulator. 
By providing posterior distributions rather than point estimates, the framework explicitly captures uncertainty arising from galactic noise. 
Moreover, once trained, the model enables rapid inference across large numbers of rangelines, making it scalable to full-planet analyses at low computational cost. 
Applications to SHARAD data show that the inferred parameter correlations are consistent with established literature and that high-probability regions correspond to physically plausible terrain properties.
Despite the accuracy of the inferred posteriors is inherently constrained by the fidelity of the simulator, the proposed SBI framework is not tied to a specific simulator, and can be applied to alternative RS instruments. 
We plan to incorporate subsurface interface depth and near-surface attenuation into the parameter space, and scale the framework to larger regions for global validation.

\small
\bibliographystyle{IEEEtranN}
\bibliography{IEEEexample}

\end{document}